\documentclass[10pt]{article}
\usepackage[utf8]{inputenc}
\usepackage[T1]{fontenc}
\usepackage{geometry}
\usepackage{multicol}
\usepackage{amsmath}
\usepackage{amsfonts}
\usepackage{graphicx}
\usepackage{url}
\usepackage{hyperref}
\usepackage[version=3]{mhchem}
\usepackage{mathtools, nccmath}
\usepackage[dvipsnames]{xcolor}
\usepackage{lipsum}
\usepackage{newpxtext,newpxmath}
\usepackage{authblk}
\usepackage{pythonhighlight}
\usepackage[font=sf,labelfont=bf]{caption}

\usepackage{xpatch}
\xpretocmd{\eqref}{Eq.~}{}{}

\usepackage[
    natbib=true,
    doi=false,
    isbn=false,
    url=false,
    eprint=false,
    maxcitenames=2,
]{biblatex}
\renewbibmacro{in:}{} 
\AtEveryBibitem{%
    \clearlist{language} 
    \clearfield{note} 
}
\DeclareNameAlias{default}{family-given} 
\title{eQuilibrator 3.0 -- a platform for the estimation of thermodynamic constants}
\author[1,2,$\dagger$]{Moritz E. Beber}
\author[3]{Mattia G. Gollub}
\author[3,4]{Dana Mozaffari}
\author[5]{Kevin M. Shebek}
\author[6,7,$\dagger$,*]{Elad Noor}

\affil[1]{\footnotesize Novo Nordisk Foundation Center for Biosustainability, Technical University of Denmark, Kemitorvet, 2800 Kongens Lyngby, Denmark}
\affil[2]{Unseen Biometrics ApS, Fruebjergvej 3, 2100 København Ø, Denmark}
\affil[3]{Department of Biosystems Science and Engineering and SIB Swiss Institute of Bioinformatics, ETH Z\"{u}rich, Basel, 4058, Switzerland}
\affil[4]{Institute of Chemical Sciences and Engineering, EPFL, Lausanne, 1015, Switzerland}
\affil[5]{Department of Chemical and Biological Engineering, Chemistry of Life Processes Institute, and Center for Synthetic Biology,, Northwestern University, Evanston, Illinois 60208, USA}
\affil[6]{Institute for Molecular Systems Biology, ETH Z\"{u}rich, Z\"{u}rich, 4093, Switzerland}
\affil[7]{Department of Plant and Environmental Sciences, Weizmann Institute of Science, Rehovot, Israel}
\affil[$\dagger$]{These authors contributed equally.}
\affil[*]{Corresponding author. Email: elad.noor@weizmann.ac.il}
\date{\today}

\newcommand{\Cmat}{\mathbf{C}}
\newcommand{\Lmat}{\mathbf{L}}
\newcommand{\Xmat}{\mathbf{X}}
\newcommand{\Smat}{\mathbf{S}}
\newcommand{\Qmat}{\mathbf{Q}}
\newcommand{\Gmat}{\mathbf{G}}
\newcommand{\Gammamat}{\mathbf{\Gamma}}
\newcommand{\Zeromat}{\mathbf{0}}
\newcommand{\Eyemat}{\mathbf{I}}
\newcommand{\PRmat}[1]{\mathbf{P}_{\mathcal{R}\left(\mathbf{#1}\right)}}
\newcommand{\PNmat}[1]{\mathbf{P}_{\mathcal{N}\left(\mathbf{#1}\right)}}
\newcommand{\Covmat}[1]{\mathbf{\Sigma}({\mathbf{#1}})}

\begin{document}

\maketitle

\begin{abstract}
eQuilibrator~\footnote{\url{equilibrator.weizmann.ac.il}} is a calculator for biochemical equilibrium constants and Gibbs free energies, originally designed as a web-based interface. While the website now counts ${\sim}1000$ distinct monthly users, its design could not accommodate larger compound databases and it lacked an application programming interface (API) for integration in other tools developed by the systems biology community. Here, we report a new python-based package for eQuilibrator, that comes with many new features such as a 50-fold larger compound database, the ability to add novel compound structures, improvements in speed and memory use, and correction for \ce{Mg^2+} ion concentrations. Moreover, it adds the ability to compute the covariance matrix of the uncertainty between estimates, for which we show the advantages and describe the application in metabolic modeling. We foresee that these improvements will make thermodynamic modeling more accessible and facilitate the integration of eQuilibrator into other software platforms.
\end{abstract}

\begin{multicols}{2}

\section{Introduction}
The field of thermodynamics started in the midst of the industrial revolution as an effort to improve mechanical engines \citep{carnot_reflexions_1824}. The phenomenal success of the theory to describe the relationships between \textit{energy}, \textit{heat}, and \textit{work} and to provide accurate predictions of what is feasible, inspired countless other scientific endeavors, including molecular dynamics and even economics \citep{jinich_quantum_2014, jinich_quantum_2018, georgescu-roegen_entropy_1999}. Curiously, thermodynamic reasoning is relatively underutilized in metabolic modeling. Reasons for this include:
\begin{itemize} 
    \item \textbf{The knowledge gap} -- equilibrium constants for most biochemical reactions have not been measured.
    \item \textbf{The computation gap} -- thermodynamic constraints tend to make metabolic models more complicated. For example, Flux Balance Analysis (FBA) with thermodynamic constraints turns from a standard Linear Problem to a Mixed-Integer one (MILP) \citep{henry_thermodynamics-based_2007,mahamkali_multitfa_2020}.
    \item \textbf{The motivation gap} -- it is not clear to everyone that using thermodynamics in models is actually necessary or even useful.
    \item \textbf{The tools gap} -- adding thermodynamics to an existing model is laborious. It involves tasks such as: mapping identifiers, adjusting the $\Delta G'^\circ$ values to the aqueous conditions, and annotating charged molecules correctly -- to name but a few.
\end{itemize}

One of the major breakthroughs in bridging the knowledge gap was achieved by \citet{lydersen_estimation_1955}, who suggested to adapt the Group Contribution method to the world of organic chemistry, as well as by \citet{joback_estimation_1987} and \citet{mavrovouniotis_group_1988} who implemented it decades later. This data-driven approach was able to cover the majority of small molecules which appear in metabolic models and obtain estimates for their Gibbs energy of formation \citep{feist_genome-scale_2007}. Since then, improvements to the accuracy and coverage of this method have been proposed \citep{jankowski_group_2008, noor_integrated_2012, noor_consistent_2013, du_temperature-dependent_2018}. Although gathering more experimentally derived equilibrium constants can still improve our estimates (in some cases, it might even be necessary), one can arguably say that the knowledge gap has been mostly addressed.

Similarly, the complexity gap has changed from a hard barrier to a minor inconvenience. Increasingly faster computers and powerful MILP solvers (such as IBM \href{https://www.ibm.com/products/ilog-cplex-optimization-studio}{CPLEX} and \href{https://www.gurobi.com/}{Gurobi} which offer free academic licenses) made it easy to solve large problems using personal computers. A task which was inconceivable only a decade ago.

The motivation gap is harder to overcome due to a chicken-and-egg problem. Since other, more technical, issues were delaying the application of thermodynamic models, it was difficult to demonstrate the usefulness of these models in reality and therefore convince the scientific community that they are worth investing in. Nevertheless, several methods which take advantage of such models already exist. Some exploit thermodynamic principles to constrain reaction directionality and metabolite concentrations \citep{holzhutter_principle_2004,henry_thermodynamics-based_2007,salvy_pytfa_2018}. Others use thermodynamic driving forces as a proxy for pathway efficiency \citep{noor_pathway_2014,hadicke_optmdfpathway:_2018}. More recently, probabilistic methods combining thermodynamic parameters have been suggested for parameter estimation \citep{lubitz_parameter_2019} and flux sampling \citep{gollub_probabilistic_2020}. These algorithms have the potential to improve the flux predictions produced by FBA \citep{noor_removing_2018}, and assist in the design of new metabolic pathways \citep{hadicke_optmdfpathway:_2018}.

It seems that the time has come to close the last remaining gap, namely the tools gap. In recent years, a plethora of software tools have facilitated the reconstruction, validation, and analysis of genome-scale metabolic models and made them a community standard which is applied in thousands of scientific projects every year 
\citep{carey_community_2020}. In 2012, the first version of a website called eQuilibrator was launched, which aimed to do the same for thermodynamic parameters \citep{flamholz_equilibratorbiochemical_2012}. eQuilibrator provides a simple search-focused interface for quickly finding a biochemical reaction's Gibbs free energy change, and is now used by $\sim$1000 distinct users every month. However, eQuilibrator was designed to be used for single reactions. Therefore, it is inefficient at querying lists of reactions and doesn't account for correlations between multiple estimates. In this paper, we present \texttt{equilibrator-api}, a new Python package which is aimed at both novice and expert programmers that want to add thermodynamic parameters to their models.

\section{Results}

\subsection{New features in eQuilibrator 3.0}

When eQuilibrator was first launched, a biochemical reaction's Gibbs free energy change $\Delta_r G'$ estimates were based on the pseudo-isomeric group contribution method \citep{noor_integrated_2012}. About two years later, we updated the back-end engine to the more recent component contribution (CC) method \citep{noor_consistent_2013}, which required developing a way to calculate CC estimates on-the-fly. In section \ref{sec:on_the_fly} we present, for the first time, the basis for these calculations. An advantage provided by the new system, which can only be accessed via the python package, is the covariance matrix for the uncertainty between estimates. In some cases, standard transformed formation ($\Delta_f G'^\circ$) or reaction ($\Delta_r G'^\circ$) energy estimates have large uncertainties when taken individually. However, uncertainties might be highly correlated, e.g., when reactions share a common compound or compounds share a common chemical group. In contrast to per-reaction and per-compound uncertainties, the covariance matrix describes the uncertainties precisely. In section \ref{sec:multivariate} we explain how the covariance matrix can be used in constraint-based thermodynamic models.

At the same time, we added a list of new features which benefit both users of the website and the Python package. These include a 50-fold larger compound database and the ability to add new compounds to a local database, support for multi-compartment reactions, changing magnesium ion concentrations, and improvements in speed and memory use. 

\subsubsection{Expanding the scope of compounds}
A frequent request from eQuilibrator users was adding compounds that are not present in the KEGG database. Therefore, we modified eQuilibrator to use MetaNetX \citep{moretti_metanetxmnxref_2021}, a database that aggregates chemicals that are relevant for metabolic models from multiple online databases, including KEGG, ChEBI, BiGG, ModelSEED, Swiss Lipids (see Figure \ref{fig:eq3_design}). This expanded the repertoire of compounds from $\sim$10,000 to $\sim$500,000, which can now be accessed using identifiers from different namespaces. As a result, eQuilibrator can be used with metabolic models from different sources (e.g., SEED or BiGG models) directly, without the need to map all compounds to KEGG identifiers in advance.

While the incorporation of more databases greatly increases the scope of eQuilibrator, many applications require the use of compounds that are still not covered. For example, metabolic pathway engineering often utilizes promiscuous enzymes to generate novel reactions, producing pathways with compounds not found in MetaNetX \cite{schwander_synthetic_2016,jeffryes_mines_2015,hadadi_atlas_2016,delepine_retropath20_2018,moriya_pathpred_2010}. To address this problem, we have extended \texttt{equilibrator-assets}, the package responsible for generating the distributed compound database, to provide methods for users to directly create new entries (see Figure \ref{fig:eq3_design}). User-specified structures (given either as InChI or SMILES) are processed with OpenBabel \cite{oboyle_open_2011} and \href{https://chemaxon.com/products/calculators-and-predictors}{ChemAxon} to generate and add compounds into the existing database. These compounds can be used directly with the \texttt{equilibrator-api} package, allowing for seamless integration of new compounds in thermodynamic analyses \cite{scheffen_new--nature_2021}.

\subsubsection{Multi-compartment reactions}

The standard calculation for reaction Gibbs energies assumes that all reactants are in the same aqueous compartment, with a constant pH, pMg, ionic strength and temperature \citep{alberty_biochemical_1994}. However, most genome-scale metabolic models describe more than one compartment, usually separated by a lipid membrane, and contain many transport reactions that span compartments with different aqueous conditions. Furthermore, the membrane between each two compartments can be associated with an electrostatic potential $\Delta \Phi$ which affects the thermodynamics of charged ions traveling between them. When a reaction involves transport of metabolite species between different compartments with different hydrogen ion activity or electrical potential, we add the following term to its $\Delta_r G'^\circ$:
\begin{equation}
    -N_H \cdot RT\ln\left( 10^{\Delta pH} \right) - Q \cdot F \Delta \Phi
\end{equation}
where $R$ is the gas constant, $T$ is the temperature, $F$ is Faraday's constant (the total electric charge of one mol of electrons -- $96.5$ kC mol$^{-1}$), $\Delta pH$ is the difference in pH between initial and final compartment, $N_H$ is the net number of hydrogen ions transported from initial to final compartment, and $Q$ is the stoichiometric coefficient of the transported charges \citep{alberty_biochemical_1994, haraldsdottir_quantitative_2012}. Note that $RT\ln\left( 10^{\Delta pH} \right)$ is the difference in potential between the compartments specifically for protons (due to the difference in their concentration), and $F \Delta \Phi$ is the difference in potential relevant to all charges (including protons).

In prior versions of eQuilibrator, adjusting the standard Gibbs energy of multi-compartment biochemical reactions, such as ones facilitated by membrane transporters, had to be performed as a post-processing step. For example, the \textit{vonBertalanffy} extension in the COBRA toolbox \citep{fleming_von_2011} performs such additional calculations as part of its pipeline. In eQuilibrator 3.0, we add functions to facilitate the adjustments required for multi-compartment reactions as part of the main package. See Section \ref{sec:mulicompartmental_example} for an example.

\subsubsection{Adjusting estimates to different pMg values}
Since ions are abundant in the cytosol and can bind metabolites to varying degrees, ion concentrations have a large effect on biochemical thermodynamics~\citep{alberty_biochemical_1994}. The concentration of protons (\ce{H+}), commonly expressed as the pH, is the most dramatic example of this phenomenon. However, this is not the only case. Magnesium ions (\ce{Mg2+}) bind to many common biochemical moieties, especially phosphate, and have been shown to play a significant role in the thermodynamics of glycolysis~\citep{vojinovic_influence_2009}. For example, the dissociation constants for \ce{ATP} and \ce{ADP} are low enough for them to be in their complex forms \ce{MgATP^2-} and \ce{MgADP-} at a physiological intracellular pMg of 3.

Every compound can be seen as an ensemble of \textit{pseudoisomers}, molecules only differing in protonation state or magnesium binding state. In a biochemical context, where all compounds are assumed to be in a buffered aqueous environment, we do not distinguish between pseudoisomers and refer to the entire ensemble as a \textit{metabolite} (note, that this assumption does not hold for transport reactions across membranes). It is thus convenient to discuss the standard \textit{transformed} Gibbs energy of formation $\Delta_f G'^\circ$ which groups together all pseudoisomers into one formation energy. It can be obtained using a Boltzmann-weighted mixture of its constituent pseudoisomers:
\begin{equation}
    \Delta_f G'^\circ = -RT\ln{\sum_j\mathrm{e}^{-\Delta_f G'^\circ(j)/RT}}\label{eq:legendre1}\,.
\end{equation}

The standard transformed Gibbs energies of formation $\Delta_f G'^\circ(j)$ for each pseudoisomer $j$ at given biochemical conditions can be calculated using the Legendre transformation
\begin{align}
    \label{eq:legendre2}
    \Delta_f G'^\circ(j) &= \Delta_f G^\circ(j) \nonumber\\
    &\color{Aquamarine} - N_H(j)[\Delta_f G^\circ(H^+) + RT\ln(10^{-pH})]\nonumber\\
    &\color{Fuchsia} -  N_{Mg}(j)[\Delta_f G^\circ(Mg^{2+}) + RT\ln(10^{-pMg})]\, \color{Black}.
\end{align}

The first term is the \textit{chemical} standard Gibbs energy of formation of the pseudoisomer. The second term describes the \textcolor{Aquamarine}{contribution of protons to the Gibbs energy as a function of the pH}. Similarly, the effect of the concentration of Mg$^{2+}$ ions (quantified as pMg) can be taken into account by adding \textcolor{Fuchsia}{a third term for the contribution of magnesium ions}. 

Affinity to \ce{Mg^2+} varies between compounds and pseudoisomers. The presence of certain chemical moieties, such as phosphate groups, tends to increase the binding affinity \citep{alberty_thermodynamics_2003}, while increasing the protonation state tends to decrease the affinity. Unfortunately, the availability of affinity constants for \ce{Mg^2+} is much lower than for \ce{H^+}. In eQuilibrator, we used $\Delta_f G'^\circ(j)$ for magnesium-bound pseudoisomers collected by \citet{vojinovic_influence_2009} and affinities predicted by \citet{du_estimating_2018}. For all other pseudoisomers, we assumed that their affinity is weak and has negligible effect on thermodynamics.

After populating the database with magnesium-bound pseudoisomers, we computed the root mean square error (RMSE) for all reactions from the NIST TECR database \citep{goldberg_thermodynamics_2004}.
When taking magnesium into account, the RMSE improved slightly from 2.99 to 2.93.

\subsubsection{Complete code refactoring}
We realized, that in order to extend this framework to new common uses-cases, such as the ones described in the previous sections, a complete refactoring was required -- creating separate modular packages for each distinct function. The original code was designed exclusively for a single use-case: starting with analyzing the chemical structures (group decompositions), reverse-transforming the measured equilibrium constants to chemical Gibbs energies \citep{alberty_legendre_1997}, solving the linear regression problems to find the group contribution energies, and using the solutions to estimate formation energies for all compounds in the KEGG database. This long procedure was not useful for users who only wanted to apply CC on a list of their own reactions.

\begin{figure*}[t!]
    \centering
    \includegraphics[width=\textwidth]{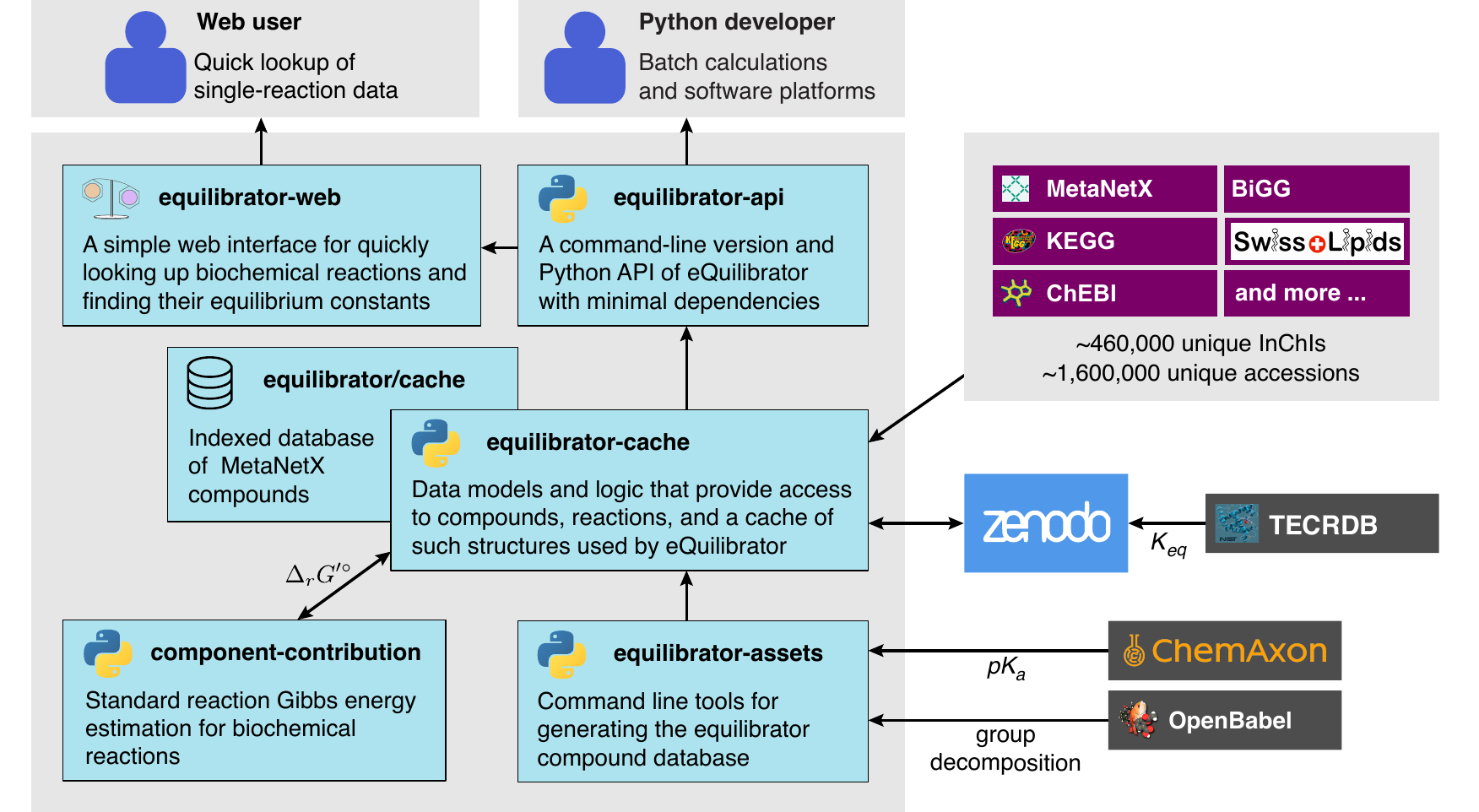}
    \caption{The design of the eQuilibrator 3.0 suite. At the core, the equilibrator-cache python package defines and manages the database of all compounds, denoted \textit{equilibrator/cache}. All compound identifiers and names come from MetaNetX, which aggregates several popular compound repositories. The component-contribution package is responsible for handling training data, group decomposition, Legendre transforms (for pH, pMg and ionic strength adjustments), and the final estimation of Gibbs energies for new compounds and reactions. Some scripts required for rebuilding the database and adding new compounds to a local database are stored in \texttt{equilibrator-assets}. All data relevant for running the eQuilibrator packages is stored in Zenodo and is freely available. This includes the experimental $K_{eq}$ data used to train the component contribution method, which comes mainly from NIST TECR database \citep{goldberg_thermodynamics_2004}. The equilibrator-api package exposes nearly all functions relevant for users with sufficient programming skills, and is required for running eQuilibrator in batch mode (e.g. on an entire metabolic model). On the other hand, the web interface provides quick and easy access to eQuilibrator estimates, but is only designed to deal with a single reaction at a time.}
    \label{fig:eq3_design}
\end{figure*}

Therefore, we have redesigned the entire component-contribution package, moved it to a new \href{https://gitlab.com/equilibrator/component-contribution}{Git repository}, and integrated it completely into a larger framework denoted \textit{eQuilibrator} (see Figure \ref{fig:eq3_design}). In addition, we raised the coding standards, e.g. by running automated tools for coverage, unit-testing, and documentation (available on \url{equilibrator.rtfd.io}). We also facilitated the installation of the packages by submitting them to the Python Package Index (\url{https://pypi.org}) and conda-forge (\url{https://conda-forge.org}).

\subsection{Fast calculation of Gibbs energies using component contributions}\label{sec:on_the_fly}

It is a specific challenge to use the Component Contribution (CC) method in a website such as eQuilibrator, since uncertainty calculations must be made on-the-fly, but the methodology for CC was developed as a one-step calculation for thousands of reactions in a model. The time needed to run CC even for a single reaction is too long to be useful for a website, even if it can be optimized and decreased to a few seconds.

Therefore, we introduced a pre-processing step in which relatively small intermediate matrices are stored in-memory and are used for fast, on-the-fly calculations. This approach provides a trade-off between memory requirements and calculation time.

The full derivation of the equations for the pre-processing vectors and matrices, required for the fast calculations, is provided in the Appendix (\ref{sec:derivation_fast_cc}). Here, we present only the final results.

\subsubsection{Pre-processing step}\label{sec:pre_processing}

In the pre-processing step, we calculate the $\boldsymbol{\upmu}$ vector, which can be seen as the formation energies of reactants and groups:
\begin{align}
    \boldsymbol{\upmu} &\equiv
    \begin{bmatrix}
		\PRmat{S} \left(\Smat^{\top}\right)^{+} + 
		\PNmat{S^\top} \Gmat \left(\Smat^{\top}\Gmat\right)^{+}
    	\\
    	\left(\Smat^{\top}\Gmat\right)^{+}
    \end{bmatrix}
    \Delta_{r}G_{obs}^{\circ}
\end{align}
where $\Smat$ is the stoichiometric matrix of the training data-set, $\Gmat$ is the group incidence matrix, and $\Delta_{r}G_{obs}^{\circ}$ are the observed chemical Gibbs energies of the training set reactions. The $()^{+}$ sign represents the matrix pseudo-inverse. The orthogonal projections $\PRmat{S}$ and $\PNmat{S^\top}$ are the projections on the range of $\Smat$ and the null-space of $\Smat^\top$ respectively. Note that orthogonal projection matrices satisfy the equations $\mathbf{P}^\top = \mathbf{P}$ and $\mathbf{P}^2 = \mathbf{P}$, and that $\PRmat{S} + \PNmat{S^\top} = \mathbf{I}$

For the covariance, we first define two parameters which represent the standard errors of the two sub-methods:
\begin{align*}
\alpha_{rc} &\equiv \frac{||\mathbf{e}_{rc}||}{\sqrt{n-\mbox{rank}(\Smat)}} \\
\alpha_{gc} &\equiv \frac{||\mathbf{e}_{gc}||}{\sqrt{n-\mbox{rank}(\Smat^{\top}\Gmat)}}\,.
\end{align*}
The numerators ($||\mathbf{e}_{rc}||$ and $||\mathbf{e}_{gc}||$) are the total residual error of both regressions. Therefore, $\alpha_{rc}$ and $\alpha_{gc}$ are the unbiased estimators of the two standard errors (i.e. the uncertainties). We also define $\alpha_\infty$ as the prior uncertainty when there is no data at all about a compound or group. It should theoretically be set to infinity, but for numerical reasons we need to choose a finite value. Choosing a value which is too high might cause large floating-point rounding errors. On the other hand, a small value would underestimate the real uncertainty. In eQuilibrator, we set $\alpha_\infty$ to $10^{5}$ kJ/mol by default, but that can easily be changed by the user.

Finally, we can define the following matrix:
\begin{align}\label{eq:L_matrix}
    \Lmat \equiv
    \begin{bmatrix}
        \alpha_{rc} ~ \Smat^{+} \PRmat{S} & \Zeromat \\
        \alpha_{gc} ~ (\Gmat^\top \Smat)^{+} \Gmat^\top \PNmat{S^\top} & \alpha_{gc} ~ (\Gmat^\top \Smat)^{+} \\
        \alpha_\infty ~ \PNmat{S^\top \Gmat} \Gmat^\top & \alpha_\infty ~ \PNmat{S^\top \Gmat}
    \end{bmatrix}^\top
\end{align}
The $\Lmat$ matrix is constructed so that $\Lmat \Lmat^\top$ is the covariance matrix of the uncertainty of $\boldsymbol{\upmu}$. For the full derivation of \eqref{eq:L_matrix}, see Appendix \ref{sec:on_the_fly_appendix}.

Although $\Lmat$ is a very wide matrix (with thousands of columns), we can greatly reduce its size by a rank revealing QR decomposition (also described in Appendix \ref{sec:on_the_fly_appendix}). We denote the reduced form by $\Lmat_q$, where $q$ is the rank of $\Lmat$ and is also equal to the number of columns in $\Lmat_q$. For the most recent eQuilibrator database $q = 669$.

\subsubsection{Estimation step}
We wish to estimate the Gibbs energies of a new set of reactions, described by a stoichiometric matrix $\bar{\Xmat}$. The top rows in this matrix correspond to known compounds that we had in the training set, while bottom rows represent new compounds. Let $\Gmat'$ be the group incidence matrix of the new compounds. The estimate for the Gibbs energies of the reactions in $\bar{\Xmat}$ will be:
\begin{equation}\label{eq:standard_dg_cc_X}
    \Delta_{r}G^{\circ}(\bar{\Xmat}) = \bar{\Xmat}^\top \begin{bmatrix} \Eyemat & \Zeromat \\ \Zeromat & \Gmat' \end{bmatrix} \boldsymbol{\upmu}
\end{equation}
and the square root of the covariance matrix will be:
\begin{align}\label{eq:uncertainty_cc_X}
\Qmat(\bar{\Xmat}) &= \bar{\Xmat}^\top \begin{bmatrix} \Eyemat & \Zeromat \\ \Zeromat & \Gmat' \end{bmatrix} \Lmat_q 
\end{align}
i.e. such that $\Covmat{\bar{X}} = \Qmat(\bar{\Xmat}) \Qmat(\bar{\Xmat})^\top$, see Appendix \ref{sec:on_the_fly_appendix} for details.

\subsubsection{Compressed storage of free energies}\label{sec:compressed}
Let us consider a system independent from eQuilibrator that maintains a very large compound database or one that requires frequent updating. Theoretically, one could pre-calculate all possible formation energies and covariances (as explained in section \ref{sec:pre_processing}) and redo that calculation every time a new compounds is added. However, this approach has two main limitations: (a) the size of the covariance matrix grows quadratically with the number of compounds, and (b) adding even one new compound requires complete recalculation of the covariance.

Here, we present for the first time a method to store a compressed version of the eQuilibrator database which can be used to generate the estimates and uncertainties using simple linear algebra (only matrix dot-products) and that grows linearly with the number of compounds both in terms of run-time and storage space.

First, we construct the $\Gmat'$ matrix for all the new compounds, i.e., a collection of all of their group vectors. The mean estimates for the formation energies are:
\begin{align}
     \bar{\boldsymbol{\upmu}} \equiv \begin{bmatrix} \Eyemat & \Zeromat \\ \Zeromat & \Gmat' \end{bmatrix} \boldsymbol{\upmu}
\end{align}

The uncertainties, however, require more attention. First, we note that only storing the uncertainty estimate of every formation energy by itself is not sufficient, since we cannot use them to calculate any off-diagonal value in the covariance matrix (see Figure \ref{fig:cov_motivation}). On the other hand, pre-calculating the full covariance matrix can require a prohibitive amount of memory. For example, for a database containing $n$ = 500,000 compounds (which is the case for MetaNetX) the covariance matrix would occupy 2 terabytes.

Fortunately, the solution provided in \eqref{eq:uncertainty_cc_X}, provides us with an opportunity. We define a new matrix:
\begin{align}
    \bar{\mathbf{L}}_q &\equiv \begin{bmatrix} \Eyemat & \Zeromat \\ \Zeromat & \Gmat' \end{bmatrix} \Lmat_q\,.
\end{align}
$\bar{\mathbf{L}}_q$ is a constant matrix of size $n$ by $q$, which amounts to only 2-3 gigabytes in our previous example. When we want to use our database to estimate the Gibbs energies of a set of reactions $\bar{\Xmat}$, the distribution of the estimates will be given by a multivariate Gaussian with mean and standard deviation:
\begin{align}
    \Delta_{r}G^{\circ} (\bar{\Xmat}) &= \bar{\Xmat}^\top \bar{\boldsymbol{\upmu}} \nonumber\\
    \Qmat(\bar{\Xmat}) &=\bar{\Xmat}^\top \bar{\mathbf{L}}_q \nonumber\\
    \Covmat{\bar{X}} &= \Qmat(\bar{\Xmat}) \Qmat(\bar{\Xmat})^\top = \bar{\Xmat}^\top \bar{\mathbf{L}}_q \bar{\mathbf{L}}_q^\top \bar{\Xmat}
\end{align}

Storing only $\bar{\boldsymbol{\upmu}}$ and $\bar{\mathbf{L}}_q$ facilitates performing accurate thermodynamic calculations for a large number of compounds without requiring extremely large amounts of memory. Additionally, the pre-processing and estimation steps are decoupled, meaning that end-users do not need the entire eQuilibrator codebase and its dependencies. Furthermore, adding new compounds to the database is straightforward and does not require updating existing entries, but rather only augmenting $\bar{\boldsymbol{\upmu}}$ and $\bar{\mathbf{L}}_q$ with the relevant data as extra rows.

\subsection{Sampling and optimization using the uncertainty covariance}\label{sec:multivariate}

\begin{figure*}[t!]
    \centering
    \includegraphics[width=\textwidth]{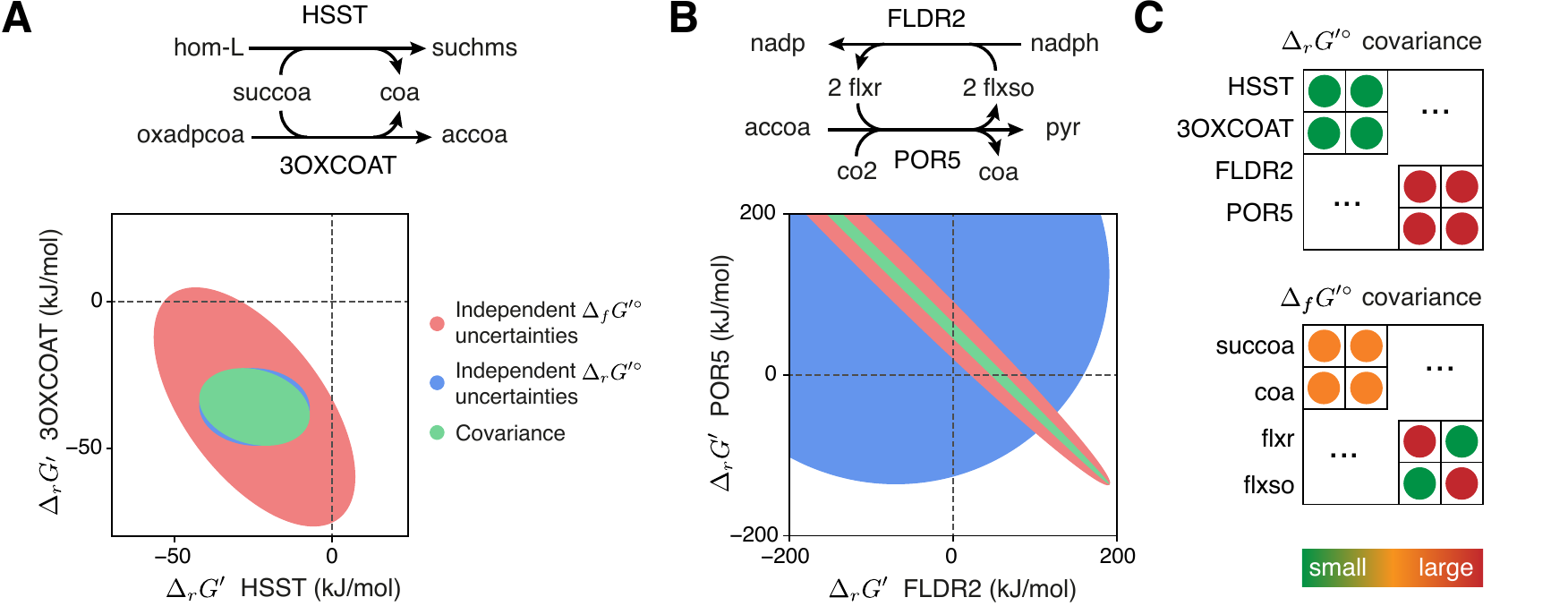}
    \caption{Examples for the importance of the covariance in the estimation uncertainty found in the iML1515 \textit{E. coli} model. (\textbf{A}) Homoserine O-succinyltransferase (HSST) and 3-oxoadipyl-CoA thiolase (3OXCOAT) both convert succinyl-CoA (succoa) to CoA. Because of the uncertainty in the $\Delta_f G'^\circ$ of CoA, computing reaction energies from independent $\Delta_f G'^\circ$ estimates results in a large uncertainty (red). As the $\Delta_f G'^\circ$ of succoa and CoA are strongly correlated, direct estimates of $\Delta_r G'^\circ$ have smaller uncertainty (blue) comparable to the uncertainty obtained using the covariance matrix of either $\Delta_f G'^\circ$ or $\Delta_r G'^\circ$ (green). (\textbf{B}) Pyruvate synthase (POR5) converts acetyl-CoA (accoa) into pyruvate (pyr) by oxidizing flavodoxin which, in iML1515, can be regenerated only through oxidation of NADPH (FLDR2). The $\Delta_r G'^\circ$ of both reactions is unknown. However, $\Delta_f G'^\circ$ of flxr and flxso must have the same value in both reactions, leading to strong correlation in the uncertainty of $\Delta_r G'^\circ$. Thus, \textit{in-vivo} synthesis of pyruvate through POR5 is unfavorable. (\textbf{C}) Covariances of $\Delta_f G'^\circ$ and $\Delta_r G'^\circ$ yield the same information about reaction energies. The small uncertainty in $\Delta_r G'^\circ$ for HSST and 3OXCOAT matches the correlation in $\Delta_f G'^\circ$ of succoa and coa, while the coupling of FLDR2 and POR5 through flavodoxin is captured by the covariance in the $\Delta_r G'^\circ$ of the two reactions. }
    \label{fig:cov_motivation}
\end{figure*}

The uncertainties of the free energies estimated with eQuilibrator are often correlated. Sometimes we have moieties whose formation energy has high uncertainty (such as coenzyme-A, Figure~\ref{fig:cov_motivation}A), but this uncertainty cancels out in reactions where the moiety is present on both sides. In contrast, there are cases where reaction energies cannot be determined because of completely uncharacterized compounds (such as flavodoxin, Figure~\ref{fig:cov_motivation}B), but using explicit formation energies reveals couplings between multiple reactions. Thus, it is often unclear whether one should use the domain of reaction energies or of formation energies. With eQuilibrator 3.0, we encourage the usage of the covariance matrix of the uncertainty when modeling multiple reactions. This matrix fully captures the correlations in the uncertainty of all quantities, and always constrains the values at least as much as when using independent uncertainties. In Figure~\ref{fig:cov_motivation}, only using the covariance matrix allows to determine reaction directions in both examples. Importantly, the covariance can be used in the domains of formation as well as reaction energies without loss of information on the reaction energies. In this section we summarize how the convariance matrix can be used in sampling and constraint based methods.

Consider a reaction network with stoichiometric matrix $\bar{\Xmat}$. The number of degrees of freedom $\bar{q}$ in the uncertainty is often smaller than the number of reactions $n$ (note that $\bar{q} \leq q = 669$). Thus, it is convenient to represent the uncertainty with a random vector $\mathbf{m} \in \mathbb{R}^{\bar{q}}$ following the standard normal distribution $\mathcal{N}(\Zeromat, \Eyemat)$ and a square root $\Qmat(\bar{\Xmat}) \in \mathbb{R}^{n \times \bar{q}}$ of the covariance $\Covmat{\bar{X}}$~\citep{gollub_probabilistic_2020}, such that:

\begin{align}
    \mathbf{m} &\sim \mathcal{N}(\Zeromat, \Eyemat) \label{eq:m_distribution} \\
    \Delta_{r}G'^{\circ} &= \Delta_{r}G'^{\circ}(\bar{\Xmat}) + \Qmat(\bar{\Xmat}) \mathbf{m} \;, \label{eq:drg_from_uncertainty}
\end{align}
where $\Eyemat$ is the $\bar{q}$-dimensional identity matrix. While $\Qmat(\bar{\Xmat})$ can be computed from the eigenvalue decomposition of $\Covmat{\bar{X}}$, this is sensitive to numerical issues if $\bar{\Xmat}$ is large. Instead, eQuilibrator computes $\Qmat(\bar{\Xmat})$ directly as described in Section \ref{sec:on_the_fly}, providing a numerically accurate result.

In order to draw random samples of the Gibbs free energies we can first draw samples of $\mathbf{m}$ using standard methods and then compute the corresponding free energies using \eqref{eq:drg_from_uncertainty}.

In a constraint-based setting, we can use the same formulation to define a quadratic constraint to bound free energies to a desired confidence level $\alpha$:
\begin{eqnarray}
    ||\mathbf{m}||_2 &\leq& \chi^2_{\bar{q};\alpha} \label{eq:quadratic_constraints}\\
    \Delta_{r}G'^{\circ} &=& \Delta_{r}G'^{\circ}(\bar{\Xmat}) + \Qmat(\bar{\Xmat}) \mathbf{m}
\end{eqnarray}
where $\chi^2_{\bar{q};\alpha}$ is the PPF (percent point function, or quantile function) of the $\chi^2$-distribution with $\bar{q}$ degrees of freedom. In Python it can be calculated using \texttt{scipy.stats.chi2.ppf()}.

When quadratic constraints cannot be used, one can replace \eqref{eq:quadratic_constraints} with upper and lower bounds for each $m_i$ separately, corresponding to a confidence interval $\alpha$ on each individual degree of freedom in the uncertainty:
\begin{align}
    |m_i| \leq \sqrt{\chi^2_{1;\alpha}} \qquad \forall \; 1\leq i\leq \bar{q}\,.
    \label{eq:indipendent_bounds}
\end{align}
Although simpler, this formulation should be used with care. Uncertainties are multivariate estimates and independent bounds can over-constrain the free energies, in particular for large networks. For example, when $\bar{q} = 50$ and $\alpha=0.95$, the bounds in \eqref{eq:indipendent_bounds} define a confidence region on $\mathbf{m}$ with an overly restrictive confidence level $\alpha^{\bar{q}}=0.08$.

\section{Discussion}
The eQuilibrator 3.0 suite marks a major shift in the focus of our software which has so far been mainly geared for single reaction searches or small biochemical networks (pathways) and exposed via a web interface. Now, we reach out to a much larger audience, including modelers who want to populate genome-scale metabolic networks with thermodynamic parameters as well as metabolic engineers that want to scan a large set of parameters for their designs. In addition, the software package can now be much more easily integrated into other python-based frameworks and pipelines such as COBRApy \citep{ebrahim_cobrapy_2013,lloyd_cobrame:_2018}, MEMOTE \citep{lieven_memote_2020}, ModelSEED \citep{seaver_modelseed_2020}, and CarveMe \citep{machado_fast_2018}. Furthermore, in section \ref{sec:compressed}, we show how one can efficiently store a set of pre-calculated matrices that can be used to calculate the final estimates (including full uncertainty matrices). This can greatly facilitate building compound databases in frameworks that are not based on Python, or that require custom optimizations and prefer not to depend on the eQuilibrator codebase directly.

The major improvements that we introduce in this work are: (1) an API supported by a refactored codebase that is much more suited to modeling applications and integration into other software, (2) improvements in speed and memory use, (3) correction for \ce{Mg^2+} ions, (4) multi-compartment reactions, (5) access to the full covariance matrix for uncertainty modeling, (6) cross-databases identification of molecules and reactions with a much larger pool of compounds (provided by MetaNetX) and the ability to add novel compounds. We continue to support community-driven development and open source standards, by publishing all the code under the permissive MIT license and making it available on \href{https://gitlab.com/equilibrator}{GitLab}. Marvin Calculator was used for estimating acid-base dissociation constants, Calculator version 18.23.0, ChemAxon (https://www.chemaxon.com), under an academic license. All other raw data needed for the algorithm is licensed under a \href{https://creativecommons.org/licenses/by/4.0/legalcode}{Creative Commons 4.0} license, and stored on \href{https://zenodo.org/}{Zenodo}.

We are open to suggestions for what could be added to eQuilibrator in future via discussions in \href{https://gitlab.com/equilibrator/equilibrator-api/-/issues}{GitLab issues} and we welcome contributions from the community. For example, new features already being considered are temperature adjustment based on separate entropy/enthalpy estimates, a fully automated script for populating metabolic SBML models with thermodynamic parameters (including multi-compartment reactions), and integration with common platforms and use-cases such as support for Thermodynamic-based Flux Analysis \citep{salvy_pytfa_2018} in COBRApy.

We believe that eQuilibrator 3.0 is a substantial step forward in closing the tools gap, and hope that together with other recent advances \citep{noor_pathway_2014,noor_protein_2016,salvy_pytfa_2018,gollub_probabilistic_2020,lubitz_parameter_2019} will bring forth the golden age of thermodynamics in the field of metabolic modeling.

\section*{Acknowledgments}

We thank Avi Flamholz for the help in writing this paper and Jörg Stelling for discussion on the effect of magnesium ions. 
This work was supported by the Swiss National Science Foundation Sinergia project \#177164. M.E.B. was partly supported by Horizon
2020 - Research and Innovation Framework Programme grant 686070 (DD-DeCaF). Kevin Shebek was supported by the U.S. Department of Energy, Office of Science, Office of Biological and Environmental Research under Award Number DE-SC0018249. 

\printbibliography[title={References}]

\section{Appendix}

\subsection{Derivation of Component Contribution pre-processing scheme}\label{sec:derivation_fast_cc}

\subsubsection{Standard Component Contribution}\label{sec:standard_cc}
According to the standard CC method \citep{noor_consistent_2013}, the vector of estimated standard Gibbs energies $\Delta_{r}G_{X}^{\circ}$ is given by the formula
\[
\Xmat^{\top} \left[ 
\underbrace{\PRmat{S} \left(\Smat^{\top}\right)^{+}}_\textrm{RC} ~+~ 
\underbrace{\PNmat{S^\top} \Gmat \left(\Smat^{\top}\Gmat\right)^{+}}_\textrm{GC}
\right] \Delta_{r}G_{obs}^{\circ}
\]
(see Section \ref{sec:on_the_fly} for the definitions of $\Xmat$, $\Smat$, $\Gmat$, $\PRmat{S}$, and $\PNmat{S}$). As indicated by the braces under the two parts, we can separate the contributions to two: Reactant Contributions (RC) and Group Contributions (GC).

The covariance matrix of the uncertainty in these Gibbs energies is given by:
\begin{align}
\Covmat{X} = \Xmat^{\top} \left[ \alpha_{rc}^2 \cdot \Cmat_{rc} + \alpha_{gc}^2 \cdot \Cmat_{gc} + \alpha_\infty^2 \cdot \Cmat_{\infty}  \right] \Xmat \label{eq:full_u}
\end{align}
where
\begin{align*}
\alpha_{rc} &\equiv \frac{||\mathbf{e}_{rc}||}{\sqrt{n-\mbox{rank}(\Smat)}} \\
\alpha_{gc} &\equiv \frac{||\mathbf{e}_{gc}||}{\sqrt{n-\mbox{rank}(\Smat^{\top}\Gmat)}} \\
\Cmat_{rc} & \equiv \PRmat{\Smat} \left(\Smat\Smat^{\top}\right)^{+} \PRmat{\Smat} \label{eq:c_rc}\\
\Cmat_{gc} & \equiv \PNmat{\Smat^\top} \Gmat \left(\Gmat^{\top}\Smat\Smat^{\top}\Gmat\right)^{+} \Gmat^{\top} \PNmat{\Smat^\top} \\
\Cmat_{\infty} & \equiv \Gmat \PNmat{\Smat^\top\Gmat} \Gmat^{\top}
\end{align*}
and $\mathbf{e}_{rc}$ and $\mathbf{e}_{gc}$ are the residuals of the reactant and group contribution regressions, and $n$ is the number of columns in $S$.

Note that the diagonal values in $\Covmat{X}$ are the \textit{squared} standard errors of the estimates of single reactions.

\subsubsection{Calculating Gibbs energy estimates on-the-fly}

What happens when we want to estimate the Gibbs energy of reactions with reactants that are not in $\Smat$? The long way would be to augment $\Smat$, $\Gmat$ and $\Xmat$ with more rows that would correspond to the new compounds. Note that if we do not have a group decomposition of one of these new compounds, there is no way to make the estimation (we cannot add ``group columns'' like we did for compounds in the training set). Fortunately, we will soon see that the effect of the added rows on the calculation is minimal, and it is easy to do the pre-processing trick we need.

Let $\Gmat'$ be the group incidence matrix of only the new compounds, and $\Xmat'$ the sub-matrix of $\Xmat$ corresponding to the new compounds. Then the new matrices we need to use for CC are:

\begin{align*}
	\bar{\Xmat} & \equiv \begin{bmatrix} \Xmat \\ \Xmat' \end{bmatrix} \\
	\bar{\Smat} & \equiv \begin{bmatrix} \Smat \\ \Zeromat \end{bmatrix} \\
	\bar{\Gmat} & \equiv \begin{bmatrix} \Gmat \\ \Gmat'\end{bmatrix}
\end{align*}
We can see that $\bar{\Smat}^\top \bar{\Gmat} = \Smat^\top \Gmat$. Since we added only zeros to $\Smat$, the range will not change, and the null-space of $\Smat^\top$ will include all the new rows. Therefore
\begin{align}
	\PRmat{\bar{\Smat}} & \equiv \begin{bmatrix}\PRmat{S} & \Zeromat \\ \Zeromat & \Zeromat \end{bmatrix}  \\
	\PNmat{\bar{\Smat}^\top} & \equiv \begin{bmatrix} \PNmat{S^\top} & \Zeromat \\ \Zeromat & \Eyemat \end{bmatrix}
\end{align}

So, the RC term will not change at all, while the GC term can be rewritten in block-matrix form:
\begin{align}\label{eq:P_N_ST_G}
\bar{\Xmat}^\top \PNmat{\bar{S}^\top} \bar{\Gmat} &=
\begin{bmatrix} \Xmat \\ \Xmat' \end{bmatrix}^\top \begin{bmatrix} \PNmat{S^\top} & \Zeromat \\ \Zeromat & \Eyemat \end{bmatrix}
\begin{bmatrix} \Gmat \\ \Gmat'\end{bmatrix} \nonumber\\
 &= \begin{bmatrix} \Xmat^\top \PNmat{S^\top} \Gmat \\ \Xmat'^\top \Gmat' \end{bmatrix}
\end{align}

Therefore, when we combine both RC and GC, the upper blocks will sum up to the same value of $\Delta_{r}G_{X}^{\circ}$ before we added $X'$, and the bottom blocks will add a new term, namely:
\begin{align}
    \Delta_{r}G_{\bar{X}}^{\circ} = \Delta_{r}G_{X}^{\circ} + \Xmat'^\top \Gmat' \left(\Smat^{\top}\Gmat\right)^{+} \Delta_{r}G_{obs}^{\circ}
\end{align}

Finally, we can define the pre-processing vectors (which depend only on the training data and not on the reaction we wish to estimate) as:
\begin{align*}
	\mathbf{v}_{r} &\equiv
	\left[
		\PRmat{S} \left(\Smat^{\top}\right)^{+} + 
		\PNmat{S^\top} \Gmat \left(\Smat^{\top}\Gmat\right)^{+}
	\right]
	\Delta_{r}G_{obs}^{\circ}
\\
	\mathbf{v}_g &\equiv \left(\Smat^{\top}\Gmat\right)^{+} \Delta_{r}G_{obs}^{\circ}
\end{align*}
and get that
\begin{eqnarray}\label{eq:fast_deltag}
\Delta_{r}G_{\bar{X}}^{\circ} &=& \Xmat^\top \mathbf{v}_r ~+~ \Xmat'^\top \Gmat' \mathbf{v}_g
\end{eqnarray}

\subsubsection{Calculating uncertainty estimates on-the-fly}\label{sec:on_the_fly_appendix}
If we look again at the definitions in section \ref{sec:standard_cc}, we can see that $\bar{\Cmat}_{rc} = \Cmat_{rc}$ is not affected by the new compounds in $X'$, besides some zero-padding for adjusting its size. For simplicity, we define the term $\Gammamat \equiv \left(\Gmat^{\top} \Smat \Smat^{\top}\Gmat\right)^{+}$, which is a symmetric matrix containing group covariances among the reactions in $\Smat$. From what we saw in the previous section, $\bar{\Smat}^\top \bar{\Gmat} = \Smat^\top \Gmat$, and therefore $\bar{\Gammamat} = \Gammamat$. We can then write:
\begin{align}
	\bar{\Cmat}_{gc} &= \PNmat{\bar{S}^\top} \bar{\Gmat} \left(\bar{\Gmat}^{\top}\bar{\Smat}\bar{\Smat}^{\top}\bar{\Gmat}\right)^{+} \bar{\Gmat}^{\top} \PNmat{\bar{S}^\top} \nonumber\\
	&= \begin{bmatrix} \PNmat{S^\top} \Gmat \\ \Gmat' \end{bmatrix}
	\Gammamat
	\begin{bmatrix} \PNmat{S^\top} \Gmat \\ \Gmat' \end{bmatrix}^\top
\nonumber\\
&=
 \begin{bmatrix} 
 \Cmat_{gc} & \PNmat{S^\top} \Gmat \Gammamat \Gmat'^\top \\ 
 \Gmat' \Gammamat \Gmat^\top \PNmat{S^\top} & \Gmat' \Gammamat \Gmat'^\top \end{bmatrix}
\end{align}
and the third term in \eqref{eq:full_u} will change to:
\begin{align}
	\bar{\Cmat}_{\infty} &=
		\begin{bmatrix} \Gmat \\ \Gmat' \end{bmatrix}
		\PNmat{S^\top\Gmat}
		\begin{bmatrix} \Gmat \\ \Gmat' \end{bmatrix}^\top
\\ &=
\begin{bmatrix}
	\Cmat_\infty &
	\Gmat \PNmat{S^\top\Gmat} \Gmat'^\top \\
	\Gmat' \PNmat{S^\top\Gmat} \Gmat^\top &
	\Gmat' \PNmat{S^\top\Gmat} \Gmat'^\top
\end{bmatrix} \nonumber
\end{align}

Finally, combining these results in the formula for the new uncertainty, we get:
\begin{align}\label{eq:fast_uncertainty}
\Covmat{\bar{X}} &= \bar{\Xmat}^{\top} \left( \alpha_{rc}\cdot \bar{\Cmat}_{rc} + \alpha_{gc}\cdot \bar{\Cmat}_{gc} + \infty\cdot \bar{\Cmat}_{\infty}  \right) \bar{X} \nonumber\\
&= 
\begin{bmatrix} \Xmat^\top & \Xmat'^\top \Gmat' \end{bmatrix}
\begin{bmatrix} \Cmat_1 & \Cmat_2 \\ \Cmat_2^\top & \Cmat_3 \end{bmatrix}
\begin{bmatrix} \Xmat \\ \Gmat'^\top \Xmat' \end{bmatrix}
\end{align}
where we define:
\begin{align*}
	\Cmat_1 &= \alpha_{rc} \cdot \Cmat_{rc} + \alpha_{gc} \cdot \Cmat_{gc} + \alpha_\infty \cdot \Cmat_\infty \\
	\Cmat_2 &= \alpha_{gc} \cdot \PNmat{S^\top} \Gmat \Gammamat + \alpha_\infty \cdot \Gmat \PNmat{S^\top\Gmat} \\
	\Cmat_3 &= \alpha_{gc} \cdot \Gammamat + \alpha_\infty \cdot \PNmat{S^\top\Gmat}\,.
\end{align*}

As can be seen in section \ref{sec:multivariate}, it is typically more convenient to use the square root of the covariance. As a Hermitian positive-definite matrix, $\Covmat{X}$ can be decomposed using the Cholesky method and therefore $\exists \mathbf{M} \in \mathbb{R}^{n \times q}$ such that $\mathbf{M}$ is lower diagonal with positive diagonal values, and $\mathbf{M} \mathbf{M}^\top = \Covmat{X}$. $q$ is the rank of $\Covmat{X}$.

In practice, however, the Cholesky decomposition often fails due to numerical issues, especially when $\Covmat{X}$ is very large. Here, we present a different approach which takes advantage of the inner structure of $\Covmat{X}$.

We start by defining the matrix $\Lmat$:
\begin{align*}
    \Lmat \equiv
    \begin{bmatrix}
        \alpha_{rc} ~ \Smat^{+} \PRmat{S} & \Zeromat \\
        \alpha_{gc} ~ (\Gmat^\top \Smat)^{+} \Gmat^\top \PNmat{S^\top} & \alpha_{gc} ~ (\Gmat^\top \Smat)^{+} \\
        \alpha_\infty ~ \PNmat{S^\top \Gmat} \Gmat^\top & \alpha_\infty ~ \PNmat{S^\top \Gmat}
    \end{bmatrix}^\top
\end{align*}
and one can convince oneself that: \[\Lmat \Lmat^\top = \begin{bmatrix} \Cmat_1 & \Cmat_2 \\ \Cmat_2^\top & \Cmat_3 \end{bmatrix}\,.\]

The problem is that $\Lmat$ is extremely rank deficient, which means that it contains a lot of linearly dependent columns (and therefore far from a compact representation of the square root). We can thus perform a rank revealing QR factorization \citep{chan_rank_1987} in order to eliminate these redundant columns and end up with a compact matrix $\Lmat_c$, such that $\Lmat_q \Lmat_q^\top = \Lmat\Lmat^\top$. This is effectively equivalent to the Cholesky decomposition of the (constant) inner part of $\Covmat{X}$.

In the case of the eQuilibrator training database, the number of columns in $\Lmat_q$ is $q = 669$, and the number of rows is $N_c + N_g$ -- or about 800. Therefore, amount of memory required to store this matrix is only about 2MB.

Note, that $\Lmat_q$ can be calculated in a pre-processing phase even before we decide on the $\bar{\Xmat}$ matrix and which new compounds are needed (represented by $\Xmat'$ and $\Gmat'$). The only post-processing step left is to define:
\begin{align}\label{eq:q_definition}
    \Qmat &= \bar{\Xmat}^\top \begin{bmatrix} \Eyemat & \Zeromat \\ \Zeromat & \Gmat' \end{bmatrix} \Lmat_q
\end{align}
and as we will see, the covariance matrix will simply be given by $\Qmat\Qmat^\top$:
\begin{align*}
     \Qmat\Qmat^\top &= \begin{bmatrix} \Xmat^\top & \Xmat'^\top \end{bmatrix} \begin{bmatrix} \Eyemat & \Zeromat \\ \Zeromat & \Gmat' \end{bmatrix} \Lmat_q \Lmat_q^\top \begin{bmatrix} \Eyemat & \Zeromat \\ \Zeromat & \Gmat'^\top \end{bmatrix} \begin{bmatrix} \Xmat \\ \Xmat' \end{bmatrix} \\
     &= \begin{bmatrix} \Xmat^\top & \Xmat'^\top \Gmat' \end{bmatrix}
\begin{bmatrix} \Cmat_1 & \Cmat_2 \\ \Cmat_2^\top & \Cmat_3 \end{bmatrix}
\begin{bmatrix} \Xmat \\ \Gmat'^\top \Xmat' \end{bmatrix} \\
    &= \Covmat{\bar{X}}
\end{align*}

\subsection{Example code for multicompartmental reactions}
\label{sec:mulicompartmental_example}

This code example below shows how to estimate $\Delta_{r}G'^{\circ}$ for glucose uptake through the phosphotransferase system. The result is $-44.8 \pm 0.6$ kJ/mol. Note that we only account for uncertainty stemming from the component-contribution method. All other estimates (e.g. based on electrostatic forces) are assumed to be precise. 

\end{multicols}
\begin{python}
from equilibrator_api import ComponentContribution, Q_

cytoplasmic_p_h = Q_(7.5)
cytoplasmic_ionic_strength = Q_("250 mM")
periplasmic_p_h = Q_(7.0)
periplasmic_ionic_strength = Q_("200 mM")
e_potential_difference = Q_("0.15 V")
cytoplasmic_reaction = "bigg.metabolite:pep = bigg.metabolite:g6p + bigg.metabolite:pyr"
periplasmic_reaction = "bigg.metabolite:glc__D = "

cc = ComponentContribution()
cc.p_h = cytoplasmic_p_h
cc.ionic_strength = cytoplasmic_ionic_strength
standard_dg_prime = cc.multicompartmental_standard_dg_prime(
    cc.parse_reaction_formula(cytoplasmic_reaction),
    cc.parse_reaction_formula(periplasmic_reaction),
    e_potential_difference=e_potential_difference,
    p_h_outer=periplasmic_p_h,
    ionic_strength_outer=periplasmic_ionic_strength,
)

print(standard_dg_prime)
\end{python}

\end{document}